\pdfoutput=1
\documentclass[preprint]{aastex62}
\usepackage{amsmath,amssymb}
\usepackage[todos]{CMImacros}

\received{XXXX}
\revised{XXXX}
\accepted{XXXX}

\shorttitle{Hyperfine line list of NH$_3$}
\shortauthors{Coles et al.}

\begin{document}
\title{Hyperfine-resolved rotation-vibration line list of ammonia (NH$_3$)}

\correspondingauthor{Andrey Yachmenev; \url{https://www.controlled-molecule-imaging.org}}
\email{andrey.yachmenev@cfel.de}

\author{Phillip A. Coles}
\affiliation{Department of Physics and Astronomy, University College London, Gower Street, WC1E 6BT London, United Kingdom}

\author[0000-0002-5167-983X]{Alec Owens}
\affiliation{Center for Free-Electron Laser Science, Deutsches Elektronen-Synchrotron DESY, Notkestrasse 85, 22607 Hamburg, Germany}
\affiliation{The Hamburg Center for Ultrafast Imaging, Universit\"at Hamburg, Luruper Chaussee 149, 22761 Hamburg, Germany}

\author[0000-0003-4395-9345]{Jochen K\"upper}
\affiliation{Center for Free-Electron Laser Science, Deutsches Elektronen-Synchrotron DESY, Notkestrasse 85, 22607 Hamburg, Germany}
\affiliation{The Hamburg Center for Ultrafast Imaging, Universit\"at Hamburg, Luruper Chaussee 149, 22761 Hamburg, Germany}
\affiliation{Department of Physics, Universit\"at Hamburg, Luruper Chaussee 149, 22761 Hamburg, Germany}

\author[0000-0001-8770-6919]{Andrey Yachmenev}
\affiliation{Center for Free-Electron Laser Science, Deutsches Elektronen-Synchrotron DESY, Notkestrasse 85, 22607 Hamburg, Germany}
\affiliation{The Hamburg Center for Ultrafast Imaging, Universit\"at Hamburg, Luruper Chaussee 149, 22761 Hamburg, Germany}

\begin{abstract}
   A comprehensive, hyperfine-resolved rotation-vibration line list for the ammonia molecule
   ($^{14}$NH$_3$) is presented. The line list, which considers hyperfine nuclear quadrupole
   coupling effects, has been computed using robust, first principles methodologies based on a
   highly accurate empirically refined potential energy surface. Transitions between levels with
   energies below $8000$~cm$^{-1}$ and total angular momentum $F\leq14$ are considered. The line
   list shows excellent agreement with a range of experimental data and will significantly assist
   future high-resolution measurements of NH$_3$, both astronomically and in the laboratory.
\end{abstract}


\section{Introduction}
\label{sec:intro}
Ammonia (NH$_3$) has been detected in a wide variety of astrophysical environments and is an
excellent molecular tracer because of its hyperfine structure. In local thermodynamic equilibrium
(LTE) conditions, the relative line strengths of the hyperfine components provide a convenient way
of deducing the optical depth~\citep{Mangum:PASP127:266}, and subsequently characterizing the
physical properties of molecular clouds~\citep{Ho:AnnRevAstronAstrophys21:239}. This approach avoids
any of the complications associated with isotopologue comparisons, such as the assumption that one
knows the isotopologue ratio, and there is no fractionation between the atomic ratio and molecular
ratio. Anomalies between the observed and theoretically predicted hyperfine spectra are frequently
observed in stellar cores, and whilst usually attributed to non-LTE
conditions~\citep{Matsakis:APJL214:L67, Stutzki:AA144:13} or systematic
infall/outflow~\citep{Park:AA376:348}, are still not well understood~\citep{Camarata:APJ806:74}.
Accounting for hyperfine effects in spectroscopic observations is thus highly desirable and a
detailed understanding of the underlying hyperfine patterns of rotation-vibration energy
levels~\citep{Twagirayezu:JCP145:144302} can even benefit the interpretation of spectra measured
with Doppler-limited resolution.

The hyperfine structure of the rovibrational energy levels is often described using effective
Hamiltonian models~\citep{Hougen:JCP57:4207, Gordy:MWMolSpec}, albeit even at 100~Hz
precision~\citep{Veldhoven:EPJD31:337}, but the limited amount of hyperfine-resolved spectroscopic
data means that these models become unreliable when extrapolating to spectral regions not sampled by
the experimental data. More successful in their predictive power over extended frequency ranges are
variational approaches, which intrinsically treat all resonant interactions between the
rovibrational states. Such calculations are becoming increasingly useful in astronomical
applications~\citep{Tennyson:MNRAS425:21,Tennyson:JMS327:73}, for example, variationally computed
molecular line lists for methane~\citep{Yurchenko:MNRAS440:1649} and
ammonia~\citep{Yurchenko:MNRAS413:1828} were used to assign lines in the near-infrared spectra of
late T dwarfs~\citep{Canty:MNRAS450:454}.

Recently, a generalized variational method for computing the nuclear quadrupole hyperfine effects in
the rovibrational spectra of polyatomic molecules was reported by two of the
authors~\citep{Yachmenev:JCP147:141101}. Utilizing this approach, we present a newly computed,
hyperfine-resolved rotation-vibration line list for $^{14}$NH$_3$ applicable for high-resolution
measurements in the microwave and near-infrared. Despite a reasonable amount of experimental and
theoretical data on the quadrupole hyperfine structure of NH$_3$ having been reported in the
literature, see \citet{Kukolich:PR156:83, Dietiker:JCP143:244305, Augustoviov:APJ824:147} and
references therein, we are aware of only two extensive, hyperfine-resolved line
lists~\citep{Coudert:AA449:855, Yachmenev:JCP147:141101}. The work presented here is an improvement
on both of these efforts and should greatly facilitate future measurements of NH$_3$, both
astronomically and in the laboratory.

The paper is structured as follows: The line list calculations are described in
\autoref{sec:calculations}, including details on the potential energy surface (PES), dipole moment
surface (DMS), electric field gradient (EFG) tensor surface, and variational nuclear motion
computations. In \autoref{sec:results}, the line list is presented along with comparisons against a
range of experimental data. Concluding remarks are offered in \autoref{sec:conc}.

\section{Line list calculations}
\label{sec:calculations}
Variational calculations employed the computer program TROVE~\citep{Yurchenko:JMS245:126,
   Yachmenev:JCP143:014105,Yurchenko:JCTC13:4368} in conjunction with a recent implementation to
treat hyperfine effects at the level of the nuclear quadrupole
coupling~\citep{Yachmenev:JCP147:141101}, which is described by the interaction of the nuclear
quadrupole moments with the electric field gradient (EFG) at the nuclei. Since the methodology of
TROVE is well documented and hyperfine-resolved calculations on the rovibrational spectrum of NH$_3$
have been described~\citep{Yachmenev:JCP147:141101}, we summarize only the key details relevant for
this work.

Initially, the spin-free rovibrational problem was solved for NH$_3$ to obtain the energies and
wavefunctions for states up to $J=14$, where $J$ is the rotational angular momentum quantum number.
The computational procedure for this stage is described in \citet{Yurchenko:MNRAS413:1828}, however,
in this work we have used a new, highly accurate, empirically refined PES~\citep{Coles:JQSRT219:119}.
For solving the pure vibrational ($J=0$) problem, the size of the primitive vibrational basis set
was truncated with the polyad number $P_{\mathrm{max}}=34$. The resulting basis of vibrational
wavefunctions was then contracted to include states with energies up to $hc\cdot20\,000$~cm$^{-1}$
($h$ is the Planck constant and $c$ is the speed of light) relative to the zero-point energy.
Multiplication with symmetry-adapted symmetric-top wavefunctions produced the final spin-free basis
set for solving the $J>0$ rovibrational problem. The final rovibrational wavefunctions, combined
with the nuclear spin functions, were used as a basis for solving the eigenvalue problem for the
total spin-rovibrational Hamiltonian. The latter is composed of a sum of the diagonal representation
of the pure rovibrational Hamiltonian and the non-diagonal matrix representation of the quadrupole
coupling. The spin-rovibrational Hamiltonian is diagonal in $F$, the quantum number of the total
angular momentum operator ${\bf F}={\bf J}+{\bf I}_{\rm N}$, which is the sum of the rovibrational
${\bf J}$ and the nuclear spin ${\bf I}_{\rm N}$ angular momentum operators.

Besides a PES, calculations require a dipole moment surface (DMS) for the computation of line
strengths, and an EFG tensor surface. The \textit{ab initio} EFG tensor surface at the quadrupolar
nucleus $^{14}$N was generated on a grid of 4700 symmetry-independent molecular geometries of NH$_3$
using the coupled cluster method, CCSD(T), with all electrons correlated in conjunction with the
augmented correlation-consistent core-valence basis set, aug-cc-pwCVQZ~\citep{Dunning:JCP90:1007,
   Kendall:JCP96:6796, Peterson:JCP117:10548}. Calculations utilized analytical coupled cluster
energy derivatives~\citep{Scuseria:JCP94:442} as implemented in the CFOUR program
package~\citep{CFOUR:2017_aa}. The elements of the EFG tensor were converted into a symmetry-adapted
form in the {\bf D}$_{\rm 3h}$(M) molecular symmetry group and represented by symmetry-adapted power
series expansions up to sixth-order. Details of the representation and least-squares fitting
procedure can be found in \citet{Yachmenev:JCP147:141101}. Similarly, the \textit{ab initio} DMS was
calculated at the CCSD(T)/aug-cc-pCVQZ level of theory with all electrons correlated on the same
grid of nuclear geometries as the EFG tensor. The least-squares fitting by analytical expansions was
performed following the method described in \citet{Yurchenko:JPCA113:11845} and
\citet{Owens:JCP148:124102}. A value of $eQ=20.44$~mb for the $^{14}$N nuclear quadrupole constant
was used in calculations~\citep{Pyykko:MolPhys106:1965}. The optimized parameters of the EFG tensor
surface along with the Fortran 90 functions to construct it are provided as supplementary material
\citep{Coles:AjP:inprep:suppmat}.

The computed hyperfine-resolved rovibrational line list for $^{14}$NH$_3$ corresponds to wavelengths
$\lambda>1.25$~\textmu{m} and considers all transitions between states with energy
$E\leq{}hc\cdot8000$~cm$^{-1}$ relative to the zero-point level and $F\leq14$, where
$F=|J-I_{\mathrm{N}}|,\ldots,J+I_\mathrm{N}$ and $I_\mathrm{N}=1$. The format of the line list
includes information on the initial and final rovibrational states involved in each transition such
as its wavenumber in cm$^{-1}$, symmetry, and quantum numbers. The line list is provided as
supplementary material \citep{Coles:AjP:inprep:suppmat}, along with programs to extract user-desired
transition data.

\section{Results}
\label{sec:results}
\begin{figure}
   \centering\includegraphics[width=0.75\linewidth]{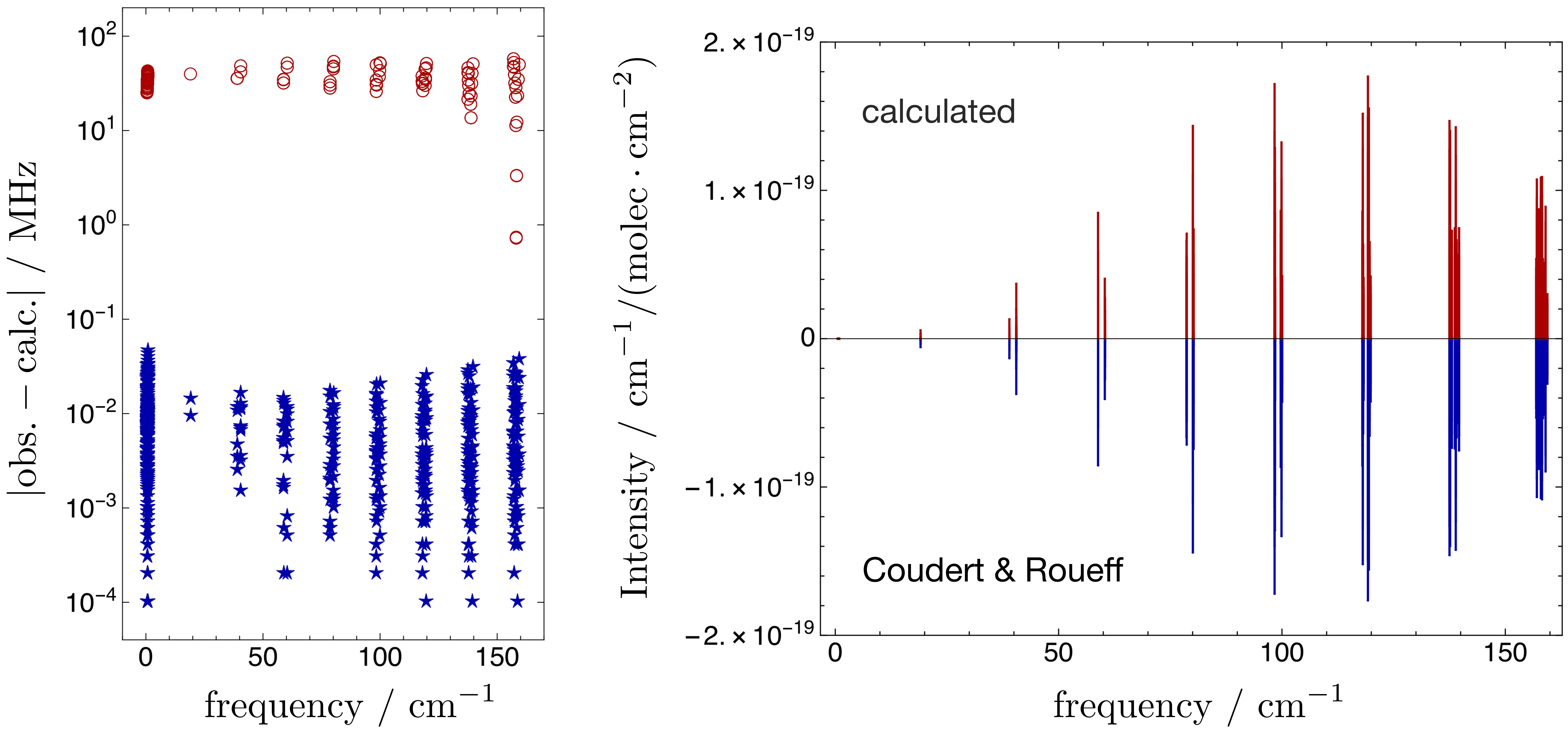}
   \caption{Discrepancies of the calculated transition frequencies of NH$_3$ relative to the
      experimental data for the ground $\nu_0$ vibrational state~\citep{Coudert:AA449:855} (left
      panel) together with the calculated and observed spectrum (right panel). The errors in the
      rovibrational frequencies are plotted with red circles while the relative errors of the
      quadrupole splittings are plotted with blue stars.}
   \label{fig:1}
\end{figure}
\begin{table}
   \centering\scriptsize%
   \caption{Discrepancies of the calculated transition frequencies (obs$-$calc) and intensities
      ($I$) with respect to the experimental data for the $\nu_2$ vibrational state of
      NH$_3$~\citep{Belov:JMolSpec189:1}. The relative errors (obs$-$calc/relative) are computed as
      quadrupole shifts with respect to the line with maximal intensity for each rovibrational band,
      i.~e., lines with the same $J'',k'',\tau_\text{inv}''$ and $J',k',\tau_\text{inv}'$, but
      different $F''$ and $F'$. The calculated relative intensities (relative $I$/calc) are obtained
      by normalizing to the maximal observed intensity (relative $I$/obs) within each rovibrational
      band.}
   \label{tab:1}
   \renewcommand{\tabcolsep}{2.5mm}
   \begin{tabular}{ccccccccrrrrrr}
\hline
$J'$ & $k'$ & $\tau_\text{inv}'$\footnote{$\tau_\text{inv}=s~\text{or}~a$ denotes \emph{symmetric} or \emph{anti-symmetric} inversion parity of the $\nu_2$ vibrational state.} & $F'$ & $J''$ & $k''$ & $\tau_\text{inv}''$ & $F''$ & \multicolumn{1}{c}{obs (MHz)} & \multicolumn{2}{c}{obs$-$calc (MHz)}
& \multicolumn{2}{c}{relative $I$} & \multicolumn{1}{c}{absolute $I$\footnote{The calculated absolute intensities for $T=300$~K are in units of cm$^{-1}$/(molecule~cm$^{-2}$)}} \\
     &      &      &      &     &     &     &     &     &  \multicolumn{1}{c}{absolute} & \multicolumn{1}{c}{relative}
& \multicolumn{1}{c}{obs} & \multicolumn{1}{c}{calc} & \\
\hline
1 & 1 & a & 1 &      2 & 1 & s & 2 &    140140.794 &       606.842 &        -0.014 &       90.00 &       90.00 &         $7.532\times 10^{-25}$ \\
1 & 1 & a & 1 &      2 & 1 & s & 1 &    140141.902 &       606.854 &        -0.003 &       30.00 &       30.00 &         $2.511\times 10^{-25}$ \\
1 & 1 & a & 2 &      2 & 1 & s & 1 &    140142.163 &       606.478 &        -0.379 &        2.00 &        1.96 &         $1.637\times 10^{-26}$ \\
1 & 1 & a & 2 &      2 & 1 & s & 3 &    140142.150 &       606.856 &         0.000 &      168.00 &      168.00 &         $1.406\times 10^{-24}$ \\
1 & 1 & a & 2 &      2 & 1 & s & 2 &    140141.427 &       606.838 &        -0.018 &       30.00 &       30.00 &         $2.511\times 10^{-25}$ \\
1 & 1 & a & 0 &      2 & 1 & s & 1 &    140143.503 &       606.862 &         0.006 &       40.00 &       40.00 &         $3.348\times 10^{-25}$ \\
2 & 2 & a & 1 &      3 & 2 & s & 2 &    741789.155 &       595.343 &         0.000 &      252.00 &      252.00 &         $1.574\times 10^{-23}$ \\
2 & 2 & a & 3 &      3 & 2 & s & 4 &    741788.397 &       595.343 &         0.000 &      540.00 &      540.00 &         $3.372\times 10^{-23}$ \\
2 & 2 & a & 3 &      3 & 2 & s & 3 &    741788.399 &       595.344 &         0.001 &       46.67 &       46.67 &         $2.914\times 10^{-24}$ \\
2 & 2 & a & 3 &      3 & 2 & s & 2 &    741788.403 &       595.349 &         0.006 &        1.33 &        1.18 &         $7.395\times 10^{-26}$ \\
2 & 2 & a & 3 &      3 & 2 & s & 4 &    741788.398 &       595.344 &         0.001 &      540.00 &      540.00 &         $3.372\times 10^{-23}$ \\
2 & 2 & a & 3 &      3 & 2 & s & 3 &    741788.388 &       595.333 &        -0.010 &       46.67 &       46.67 &         $2.914\times 10^{-24}$ \\
2 & 2 & a & 3 &      3 & 2 & s & 2 &    741788.355 &       595.301 &        -0.042 &        1.33 &        1.18 &         $7.395\times 10^{-26}$ \\
2 & 2 & a & 2 &      3 & 2 & s & 3 &    741787.015 &       595.324 &        -0.018 &      373.33 &      373.33 &         $2.331\times 10^{-23}$ \\
2 & 2 & a & 2 &      3 & 2 & s & 2 &    741787.020 &       595.330 &        -0.012 &       46.67 &       46.67 &         $2.914\times 10^{-24}$ \\
2 & 2 & a & 2 &      3 & 2 & s & 3 &    741787.019 &       595.328 &        -0.014 &      373.33 &      373.33 &         $2.331\times 10^{-23}$ \\
2 & 2 & a & 2 &      3 & 2 & s & 2 &    741786.987 &       595.297 &        -0.045 &       46.67 &       46.67 &         $2.914\times 10^{-24}$ \\
2 & 0 & a & 3 &      3 & 0 & s & 4 &    769710.287 &       576.907 &         0.000 &      540.00 &      540.00 &         $1.199\times 10^{-22}$ \\
2 & 0 & a & 2 &      3 & 0 & s & 3 &    769710.281 &       576.932 &         0.026 &      373.33 &      373.35 &         $8.291\times 10^{-23}$ \\
2 & 0 & a & 3 &      3 & 0 & s & 4 &    769710.289 &       576.909 &         0.002 &      540.00 &      540.00 &         $1.199\times 10^{-22}$ \\
2 & 0 & a & 2 &      3 & 0 & s & 3 &    769710.277 &       576.928 &         0.022 &      373.33 &      373.35 &         $8.291\times 10^{-23}$ \\
2 & 0 & a & 1 &      3 & 0 & s & 2 &    769710.000 &       576.890 &        -0.017 &      252.00 &      252.00 &         $5.596\times 10^{-23}$ \\
2 & 0 & a & 3 &      3 & 0 & s & 3 &    769708.896 &       576.915 &         0.009 &       46.67 &       46.67 &         $1.036\times 10^{-23}$ \\
2 & 0 & a & 3 &      3 & 0 & s & 2 &    769710.630 &       576.760 &        -0.147 &        1.33 &        1.18 &         $2.618\times 10^{-25}$ \\
2 & 0 & a & 2 &      3 & 0 & s & 2 &    769712.123 &       576.885 &        -0.022 &       46.67 &       46.67 &         $1.036\times 10^{-23}$ \\
2 & 1 & a & 3 &      3 & 1 & s & 3 &    762851.494 &       590.129 &        -0.037 &       46.67 &       21.78 &         $4.685\times 10^{-24}$ \\
2 & 1 & a & 3 &      3 & 1 & s & 4 &    762852.624 &       590.166 &         0.000 &      252.00 &      252.00 &         $5.421\times 10^{-23}$ \\
2 & 1 & a & 1 &      3 & 1 & s & 2 &    762852.624 &       590.163 &        -0.003 &      540.00 &      117.60 &         $2.530\times 10^{-23}$ \\
2 & 1 & a & 3 &      3 & 1 & s & 2 &    762852.942 &       590.102 &        -0.064 &        1.33 &        0.55 &         $1.185\times 10^{-25}$ \\
2 & 1 & a & 2 &      3 & 1 & s & 3 &    762852.209 &       590.160 &        -0.006 &      373.33 &      174.22 &         $3.748\times 10^{-23}$ \\
2 & 1 & a & 2 &      3 & 1 & s & 2 &    762853.684 &       590.160 &        -0.006 &       46.67 &       21.78 &         $4.685\times 10^{-24}$ \\
1 & 0 & s & 0 &      0 & 0 & a & 1 &    466243.620 &      -610.769 &        -0.007 &        4.00 &        4.31 &         $5.656\times 10^{-24}$ \\
1 & 0 & s & 2 &      0 & 0 & a & 1 &    466245.605 &      -610.762 &         0.000 &       20.00 &       20.00 &         $2.625\times 10^{-23}$ \\
1 & 0 & s & 1 &      0 & 0 & a & 1 &    466246.945 &      -610.740 &         0.022 &       12.00 &       12.93 &         $1.697\times 10^{-23}$ \\
\hline
\end{tabular}
\end{table}
\begin{table}
   \centering\scriptsize%
   \caption{Discrepancies of the calculated transition frequencies (obs$-$calc) with respect to the
      experimental data for the $\nu_1$, $\nu_3^{\pm1}$, $2\nu_4^0$, $2\nu_4^{\pm2}$ bands of
      NH$_3$~\citep{Dietiker:JCP143:244305}. The relative errors (obs$-$calc/relative) are computed
      as quadrupole shifts with respect to the line with maximal intensity for each rovibrational
      band, i.~e., lines with the same $J'',k'',\tau_\text{inv}''$ and $J',k',\tau_\text{inv}'$, but
      different $F''$ and $F'$.}
   \label{tab:2}
\renewcommand{\tabcolsep}{2.5mm}
\begin{tabular}{cccccccccrrrr}
\hline
Vibr. level & $J'$ & $k'$ & $\tau_\text{inv}'$\footnote{$\tau_\text{inv}=s~\text{or}~a$ denotes \emph{symmetric} or \emph{anti-symmetric} inversion parity of the vibrational state.} & $F'$ & $J''$ & $k''$ & $\tau_\text{inv}''$ & $F''$ & \multicolumn{1}{c}{obs (MHz)} & \multicolumn{2}{c}{obs$-$calc (MHz)} & Intensity\footnote{The calculated absolute intensities for $T=300$~K are in units of cm$^{-1}$/(molecule~cm$^{-2}$)} \\
& & & & & & & & & & absolute & relative \\
\hline
$\nu_1$ &  1 & 1 & a & 1 &      2 & 1 & s & 2 &  101166502.618 &      -115.842 &        -0.009 &               $1.228\times 10^{-21}$ \\
        &  1 & 1 & a & 2 &      2 & 1 & s & 2 &  101166503.217 &      -115.850 &        -0.016 &               $4.094\times 10^{-22}$ \\
        &  1 & 1 & a & 1 &      2 & 1 & s & 1 &  101166503.787 &      -115.673 &         0.160 &               $4.094\times 10^{-22}$ \\
        &  1 & 1 & a & 2 &      2 & 1 & s & 3 &  101166503.877 &      -115.833 &         0.000 &               $2.292\times 10^{-21}$ \\
        &  1 & 1 & a & 0 &      2 & 1 & s & 1 &  101166505.196 &      -115.782 &         0.051 &               $5.458\times 10^{-22}$ \\
        &  0 & 0 & a & 1 &      1 & 0 & s & 1 &  100580586.109 &       -76.131 &        -0.029 &               $2.249\times 10^{-21}$ \\
        &  0 & 0 & a & 1 &      1 & 0 & s & 2 &  100580587.338 &       -76.102 &         0.000 &               $3.748\times 10^{-21}$ \\
        &  0 & 0 & a & 1 &      1 & 0 & s & 0 &  100580589.137 &       -76.103 &        -0.001 &               $7.704\times 10^{-29}$ \\
\hline
$\nu_3^{\pm 1}$ & 0 & 0 & a & 1 &     1 & 1 & a & 0 &  103686651.285 &      -279.126 &        -0.136 &         $2.231\times 10^{-29}$ \\
                & 0 & 0 & a & 1 &     1 & 1 & a & 2 &  103686652.364 &      -278.989 &         0.000 &         $1.784\times 10^{-21}$ \\
                & 0 & 0 & a & 1 &     1 & 1 & a & 1 &  103686652.874 &      -279.109 &        -0.119 &         $1.070\times 10^{-21}$ \\
\hline
$2\nu_4^0$ &  0 & 0 & a & 1 &     1 & 0 & s & 1 &  97010274.474 &       148.905 &         0.164 &             $1.609\times 10^{-22}$ \\
           &  0 & 0 & a & 1 &     1 & 0 & s & 0 &  97010277.532 &       148.929 &         0.188 &             $9.758\times 10^{-30}$ \\
           &  0 & 0 & a & 1 &     1 & 0 & s & 2 &  97010275.524 &       148.741 &         0.000 &             $2.681\times 10^{-22}$ \\
\hline
$2\nu_4^{\pm 2}$ & 0 & 0 & a & 1 &     1 & 1 & a & 0 &  97485902.014 &       670.652 &        -0.021 &            $1.194\times 10^{-29}$ \\
                 & 0 & 0 & a & 1 &     1 & 1 & a & 2 &  97485902.944 &       670.672 &         0.000 &            $3.540\times 10^{-22}$ \\
                 & 0 & 0 & a & 1 &     1 & 1 & a & 1 &  97485903.543 &       670.666 &        -0.006 &            $2.124\times 10^{-22}$ \\
\hline
\end{tabular}
\end{table}
In \autoref{fig:1}, \autoref{tab:1} and \autoref{tab:2}, the predicted quadrupole hyperfine
transition frequencies and intensities for NH$_3$ are compared with the available experimental data
for the rotational transitions in the ground vibrational~\citep{Coudert:AA449:855} and
$\nu_2$~\citep{Belov:JMolSpec189:1} states, and rovibrational transitions from the ground to the
$\nu_1$, $\nu_3^{\pm1}$, $2\nu_4^0$, and $2\nu_4^{\pm2}$ vibrational
states~\citep{Dietiker:JCP143:244305}. A detailed survey of the available experimental and
theoretical data for the quadrupole hyperfine structure of NH$_3$ can be found in
\citet{Dietiker:JCP143:244305} and \citet{ Augustoviov:APJ824:147}.

The absolute errors in the rovibrational frequencies are within the accuracy of the underlying
PES~\citep{Coles:JQSRT219:119} and are reflective of what is achievable with variational nuclear
motion calculations, i.~e., sub-cm$^{-1}$ or better. To estimate the accuracy of the predicted
quadrupole splittings and the underlying EFG surface, we have subtracted the respective error in the
rovibrational frequency unperturbed from the quadrupole interaction effect for each transition. The
resulting errors range from 0.1 to 46~kHz for the ground vibrational state (\autoref{fig:1}) and
from 1 to 64 kHz for the $\nu_2$ state (\autoref{tab:1}). Notably, two lines in \autoref{tab:1} have
inconsistently large deviations of 379 and 147 kHz from experiment, which do not correlate with the
systematic errors of the calculation, but these lines have very small intensities and may have been
misassigned. The root-mean-square errors for the ground vibrational and $\nu_2$ states are 9~kHz and
72~kHz, or 20~kHz if neglecting the two lines with irregular deviations, respectively. For other
fundamental and overtone bands listed in \autoref{tab:2}, the discrepancies are larger by up to
160~kHz, however, the estimated uncertainty of the experimental data is
$\pm100$~kHz~\citep{Dietiker:JCP143:244305}, giving us confidence that the errors in our predictions
are reasonably consistent. Overall, the agreement of the presented line list with experiment has
improved in comparison to the previous theoretical study~\citep{Yachmenev:JCP147:141101}, which was
based on an older PES~\citep{Yurchenko:JMolSpec268:123} and an EFG tensor computed with a
lower-level of \emph{ab initio} theory.

\begin{figure}
   \includegraphics[width=\linewidth]{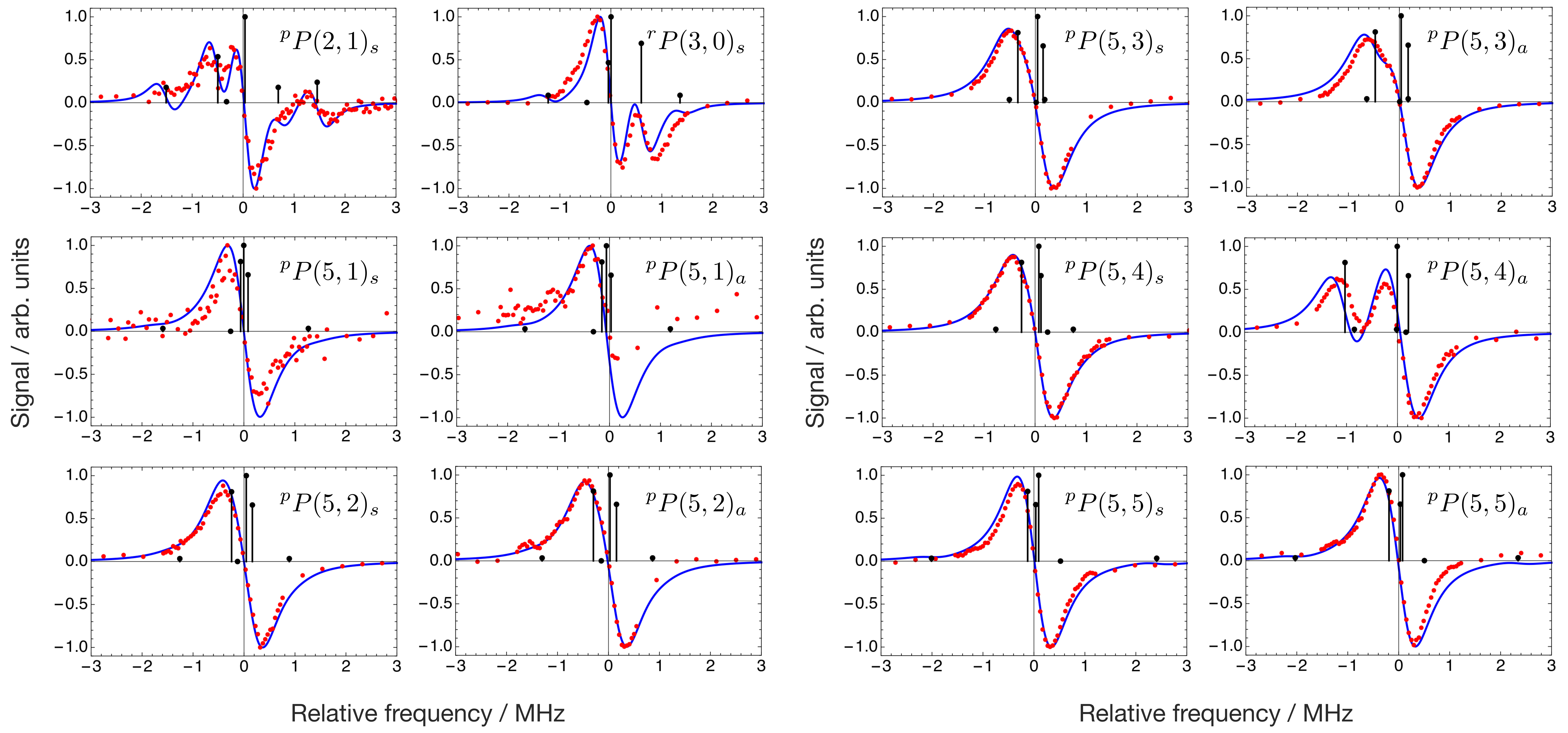}
   \caption{Comparison of the calculated (blue line) and observed (red
      dots)~\citep{Twagirayezu:JCP145:144302} saturation dip line shapes for the
      $^{\Delta K_a}\Delta J(J'',K_a'')_{\tau_\text{inv}''}$ transitions of the $\nu_1+\nu_3$ band
      of NH$_3$ ($\tau_\text{inv}''=s~\text{or}~a$ denotes \emph{symmetric} or \emph{anti-symmetric}
      inversion parity of the ground vibrational state, and $''$ denotes the lower state). Black
      stems depict the calculated stick spectrum. The experimental and calculated intensities are
      normalized to the respective maximal values. The measured (calculated) zero-crossing
      wavenumbers, in cm$^{-1}$, are 6572.85349 (6572.81120) for $^pP(2,1)_s$, 6544.32154
      (6544.28589) for $^rP(3,0)_s$, 6513.77250 (6513.73344) for $^pP(5,1)_s$, 6513.65575
      (6513.59558) for $^pP(5,1)_a$, 6521.97101 (6521.93627) for $^pP(5,2)_s$, 6522.23374
      (6522.25564) for $^pP(5,2)_a$, 6529.18969 (6529.18151) for $^pP(5,3)_s$, 6528.76857
      (6528.74384) for $^pP(5,3)_a$, 6536.59280 (6536.55360) for $^pP(5,4)_s$, 6537.68063
      (6538.22791) for $^pP(5,4)_a$, 6542.62402 (6542.63130) for $^pP(5,5)_s$, 6542.42400
      (6542.44221) for $^pP(5,5)_a$.}
   \label{fig:2}
\end{figure}
\autoref{fig:2} shows comparisons with the sub-Doppler saturation dip spectroscopic measurements for
the $\nu_1+\nu_3$ band of NH$_3$~\citep{Twagirayezu:JCP145:144302,Sears:privcomm:2017}. Saturation
dip line shapes were calculated as the intensity-weighted sums of Lorentzian line shape
derivatives~\citep{Axner:JQSRT68:299} with a half-width-at-half-maximum (HWHM) of the absorption
profile of 290~kHz and a HWHM-amplitude of the experimentally applied frequency-modulation dither of
150~kHz~\citep{Sears:privcomm:2017}. A slightly larger HWHM was employed for the measured
$^pP(5,K_a'')$ transitions~\citep{Sears:privcomm:2017} and we have found a value of 500~kHz
reproduces these line shapes well. Overall, the computed saturation dip profiles are in excellent
agreement with experiment. Notably, in our previous theoretical
study~\citep{Yachmenev:JCP147:141101} we could not explain the observed double peak feature of the
$^pP(5,4)_a$ transition and instead predicted a double peak structure in the $^pP(5,3)_a$ transition
not seen in the experimental profile. This has now been rectified due to the use of a much improved
and more reliable PES~\citep{Coles:JQSRT219:119} and the consideration of core-valence electron
correlation in the calculation of the EFG tensor surface. Interestingly, the observed splitting in
the $^pP(5,4)_a$ transition at 6777.63638~cm$^{-1}$ arises because the upper state is in fact a
superposition of the three states $|J,k,\tau\rangle=|4,2,0\rangle$ of $\nu_3+\nu_4+3\nu_2$,
$|4,3,1\rangle$ of $(\nu_1+\nu_3)^-$ and $|4,3,0\rangle$ of $(\nu_1+\nu_3)^-$ with approximately
equal squared-coefficient contributions, where $\tau$ reflects the rotational parity defined as
$(-1)^\tau$.

\section{Conclusions}
\label{sec:conc}
A new rotation-vibration line list for $^{14}$NH$_3$, which accounts for nuclear quadrupole
hyperfine effects, has been presented. Comparisons with a range of experimental results showed
excellent agreement, validating the computational approach taken. Notably, the new line list allowed
to resolve line-shape discrepancies when compared with a previous hyperfine-resolved line list
computed by two of the authors~\citep{Yachmenev:JCP147:141101}. Due to the variational approach
taken, such improvements can be expected across the $0$--$8000$~cm$^{-1}$ region and can be
attributed to the use of a highly accurate empirically refined PES and more rigorous electronic
structure calculations for the EFG tensor surface. The line list contains detailed information,
e.~g., symmetry and quantum number labeling, for each transition, which will be extremely useful for
future analysis of hyperfine-resolved ammonia spectra. Natural extensions to our calculations would
be the consideration of a larger wavenumber range and higher energy level threshold, however, work
in this direction will only be undertaken if there is a demand for such data.

The spectrum of ammonia is also of interest regarding a possible temporal or spatial variation of
the proton-to-electron mass ratio $\mu$~\citep{Veldhoven:EPJD31:337}. If any such variation has
occurred, it would manifest as tiny but observable shifts in the frequencies of certain transitions.
Constraints on a varying $\mu$ have been deduced using NH$_3$ in our
Galaxy~\citep{Levshakov:AA524:A32}, in objects at high-redshift, for example, the system B0218$+$357
at redshift $z\sim0.685$~\citep{Flambaum:PRL98:240801, Murphy:Science320:1611, Kanekar:ApJL728:L12}
or PKS1830$-$211 at $z\sim0.886$~\citep{Henkel:AA500:725}, and are possible in high-precision
laboratory setups~\citep{Veldhoven:EPJD31:337, Cheng:PRL117:253201}. Studying the mass
sensitivity~\citep{Owens:MNRAS450:3191, Owens:PRA93:052506} of the hyperfine transitions could
reveal promising spectral regions to guide future measurements of ammonia, ultimately leading to
tighter constraints on drifting fundamental constants.

\acknowledgments%
We gratefully acknowledge Trevor Sears for providing us with their original experimental data.

This work has been supported by the \emph{Deutsche Forschungsgemeinschaft} (DFG) through the
excellence cluster ``The Hamburg Center for Ultrafast Imaging -- Structure, Dynamics and Control of
Matter at the Atomic Scale'' (CUI, EXC1074), by the Helmholtz Association ``Initiative and
Networking Fund'', and by the COST action MOLIM (CM1405). A.~O.\ gratefully acknowledges a
fellowship from the Alexander von Humboldt Foundation.

\bibliographystyle{aasjournal}

\end{document}